%% file: main.tex
\documentclass[conference]{IEEEtran}
\usepackage[utf8]{inputenc}
\usepackage{amsmath,amsthm,mathrsfs}
\usepackage{textgreek}
\usepackage[english]{babel}
\usepackage{textcomp}
\usepackage{graphicx,wrapfig}
\usepackage{amssymb}

\usepackage{calc}
\usepackage{rotating}
\usepackage[usenames,dvipsnames]{color}
\usepackage{fancyhdr}
\usepackage{longtable}
\usepackage{ifluatex}
\ifluatex
  \usepackage{pdftexcmds}
  \makeatletter
  \let\pdfstrcmp\pdf@strcmp
  \let\pdffilemoddate\pdf@filemoddate
  \makeatother
\fi
\usepackage{rotating}
\usepackage[usenames,dvipsnames]{color}
\usepackage{fancyhdr}
\usepackage[english]{babel}
\usepackage[backend=bibtex,
style=numeric,
bibencoding=ascii,
sorting=none
]{biblatex}

\usepackage{pifont}
\usepackage{xcolor}
\usepackage{dirtytalk}
\usepackage{hyperref}
\hypersetup{
    colorlinks,
    linkcolor={red!50!black},
    citecolor={blue!50!black},
    urlcolor={blue!80!black}
}

\begin{document}
%
\title{Modeling and Control of an Autonomous Three Wheeled Mobile Robot with Front Steer}

\author{\IEEEauthorblockN{Ayush Pandey, Siddharth Jha and Debashish Chakravarty}
\IEEEauthorblockA{Autonomous Ground Vehicle Research Group\\
Indian Institute of Technology (IIT)\\
Kharagpur, West Bengal, India - 721302\\
Email: ayushpandey@iitkgp.ac.in, thesidjway@iitkgp.ac.in}}


%


\maketitle

\begin{abstract}
Modeling and control strategies for a design of an autonomous three wheeled mobile robot with front wheel steer is presented. Although, the three-wheel vehicle design with front wheel steer is common in automotive vehicles used often in public transport, but its advantages in navigation and localization of autonomous vehicles is seldom utilized. We present the system model for such a robotic vehicle. A PID controller for speed control is designed for the model obtained and has been implemented in a digital control framework. The trajectory control framework, which is a challenging task for such a three-wheeled robot has also been presented in the paper. The derived system model has been verified using experimental results obtained for the robot vehicle design. Controller performance and robustness issues have also been discussed briefly.
\end{abstract}


%
\IEEEpeerreviewmaketitle

\section{Introduction}
With the rise in research and development of autonomous robots in the past decade, there has been an increased focus on control strategies for the robots to achieve robust and optimal performance. A clear application of the research on autonomous robots is the self-driving car, which has already started to change the commute in many cities all around the world. In the coming years, we are bound to discover more such self-driving vehicles on the roads and not just cars. An example is the research on three-wheeled self-driving trikes \cite{tyler} with an aerodynamic design which could effectively be deployed in the future for shared public transport. Similar is the design of a passively stabilized bicycle \cite{own}. Mercedes-Benz are working on an electric vehicle \cite{mercedes} with a related mechanical design. All of these designs are common in the sense that they are front steered and are equaivalent to a three wheel vehicle design. Control and stability are major challenges with such three wheeled vehicles. \cite{jignesh} and references their in give an account of the study of stability of three wheeled vehicles. This paper focuses on the control aspects for such a vehicle design. \\
There are certain distinct advantages that can be had with a three-wheeled robot design. The steer using the front wheel is quite close in working to the design of cars. However, the localization and navigation of such three wheeled vehicles is completely different. If the drive actuation to the vehicle is also provided in the front wheel, as is the case for our robot design, the two rear wheels are free. These two wheels can be very effectively used for accurate localization, which would have been otherwise impossible in a rear wheel driven vehicle. The absence of actuators in the wheels gives way to precise localization which in turn helps in better trajectory following and navigation of the vehicle. Although, the modified mechanical design has advantages in navigation, it poses challenges in modeling and control strategies which haven't been discussed in the existing literature concerned with autonomous robots. This paper aims to cover this gap by identifying the model and control design of a three wheeled mobile robot with front steer and front wheel drive.

\subsection{Background}
There has been extensive research on low level control of autonomous mobile robots (\cite{robot1}, \cite{robot2}). Low-level control strategies for mobile robots (autonomous or otherwise) are heavily dependent on the dynamics of the robot. Most common mobile robots today are based on the differential drive model, in which two powered wheels are used to both drive the robot and change its direction. The research on control of such robot vehicles is vast and is not of concern in this paper. However, it is of importance to note that the control strategies for a differential drive robot are completely different and do not apply to other robot designs such as omnidirectional mobile robots \cite{robot3} or ackerman drive robots \cite{robot4} which are very similar to modern cars. This paper presents a low-level control model for an new kind of steering geometry, consisting of a three wheeled robot which is both steered and powered using the front wheel. This type of steering geometry has several advantages (as described later in detail) for the purpose of localization and motion planning. Similar kind of designs have been discussed in \cite{robot5} and \cite{robot6} but the work on modeling and control of such robot designs is still in a nascent stage. \\

\section{Objectives}
\begin{figure}[htbp]
 			  \centering
 			  \def\svgscale{0.25}
  			  \tiny{
  			  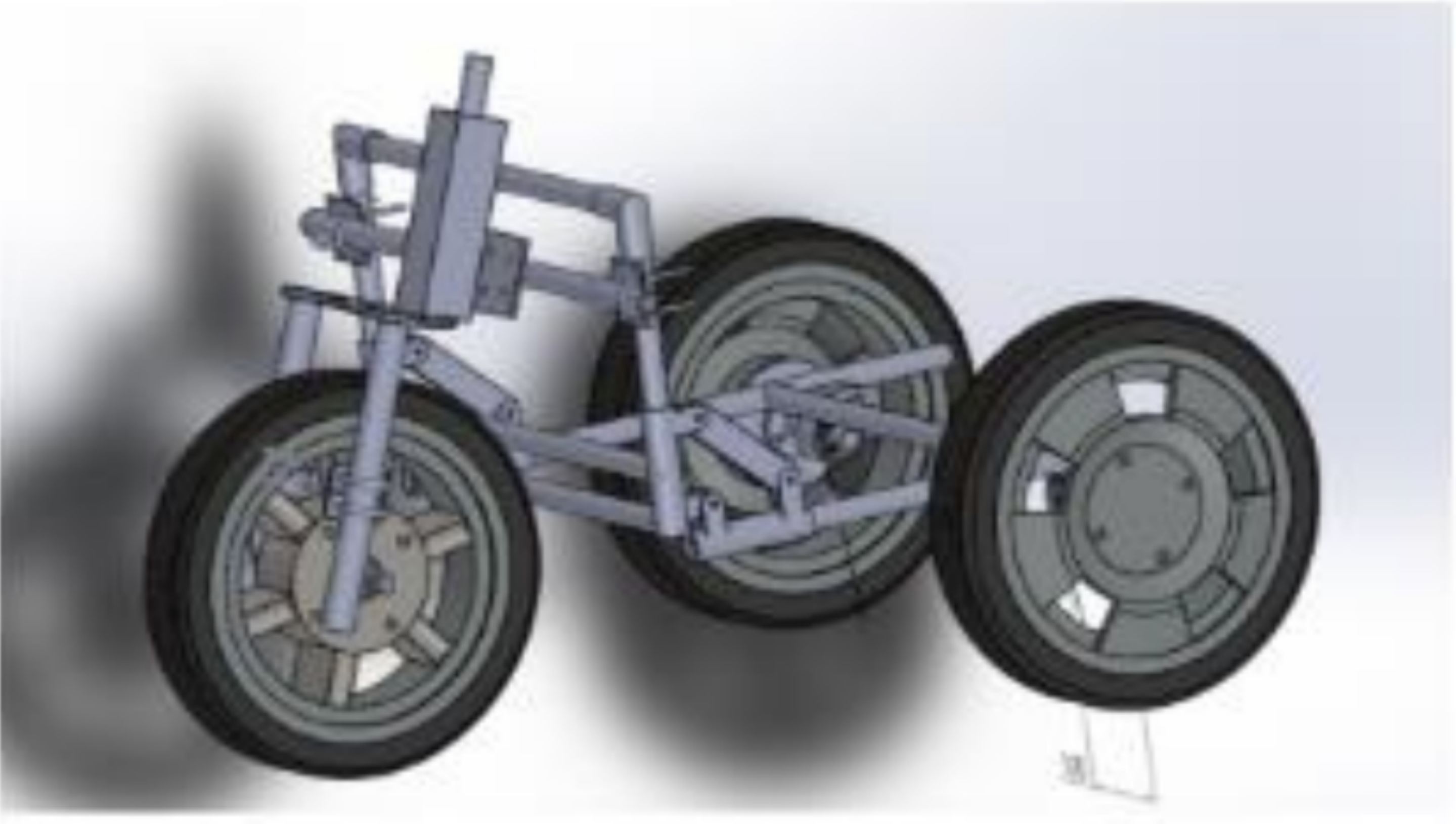}
 			  \caption{The three wheeled autonomous mobile robot design with front steer}
 			 \label{robot}
 		\end{figure}
For the three wheeled autonomous mobile robot as shown in Fig.(\ref{robot}), we aimed to design the complete low-level control system. There are three main coupled subsystems working in the low-level control of the robot viz. velocity control, steering control and trajectory control. To design appropriate control algorithms, we first aimed to identify the system model and validate it with experimental results. Our main focus in this paper is on velocity control system, however, using similar methods we propose steering control as well. The trajectory control for this robot design is a challenging task because of the uniqueness of the mechanical design. Towards the end of this paper, we have proposed a trajectory control methodology and experimental results for the same as well. 
\subsection{Velocity Control}

\begin{figure}[htbp]
 			  \centering
 			  \def\svgscale{0.25}
  			  \tiny{
  			  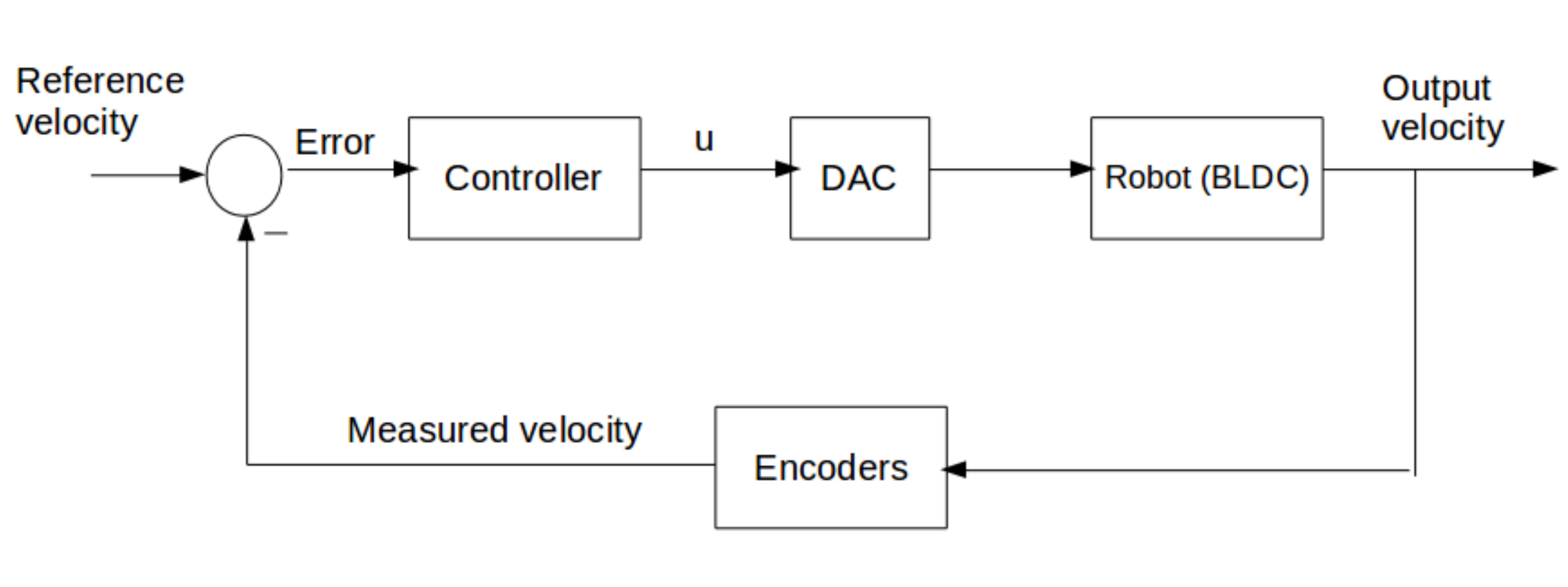}
 			  \caption{Velocity Control System Block Diagram}
 			 \label{veloblock}
 		\end{figure}
The front wheel of the robot drives the robot using a brushless DC motor which provides the required thrust. The BLDC is in an outer closed loop control as shown in the velocity control system block diagram in Fig.(\ref{veloblock}). We aimed to design a controller which achieves optimal performance for a step input. Approximating the model for the robot in velocity control by the BLDC model, we first aimed to identify the plant model and then proceeded towards control design. The controller implementation and experimental performance analysis have also been considered in the paper.
\subsection{Trajectory and Steering Control}
\begin{figure}[htbp]

			  \centering
			  \def\svgscale{0.25}
			  \tiny{
			  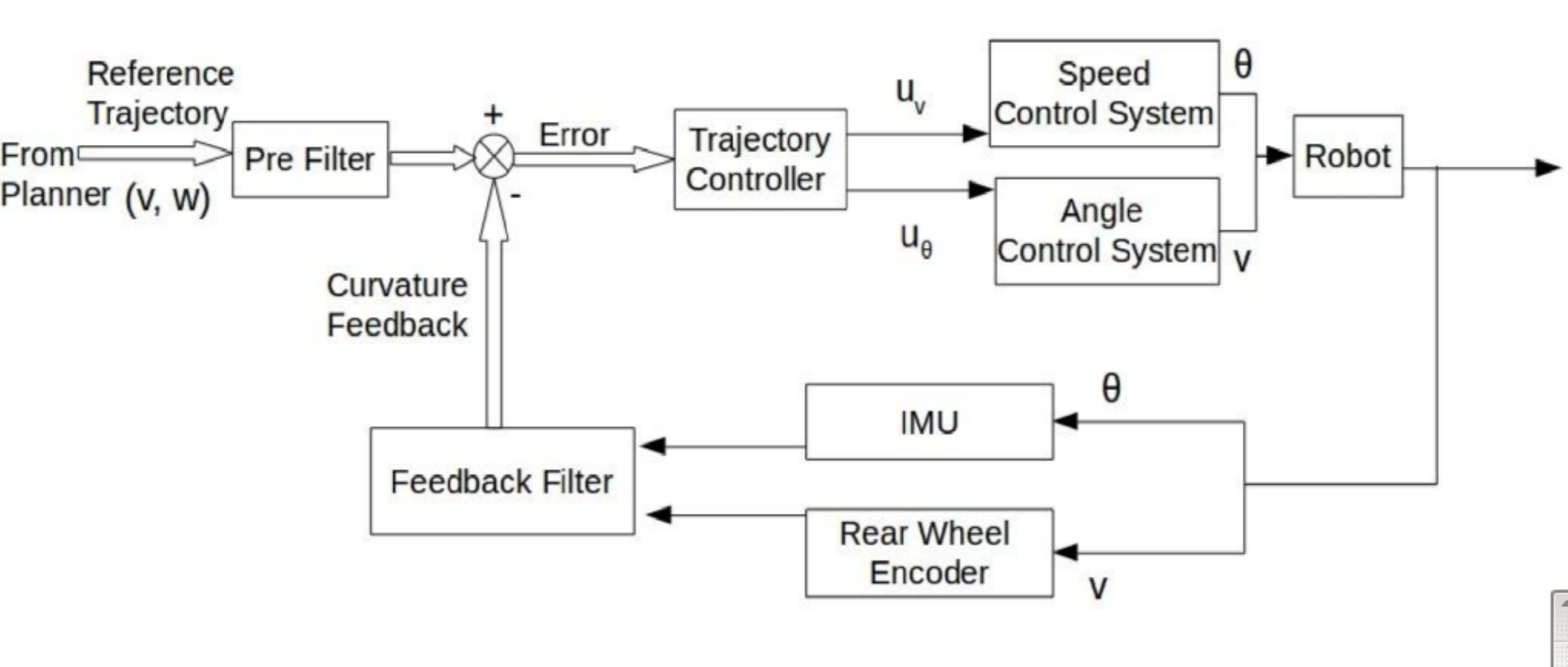}
			  \caption{Trajectory Control System Block Diagram}
			 \label{trablock}
		\end{figure}

A large class of control problems consist of planning and following a trajectory in the presence of noise and uncertainty \cite{murray}. Trajectories become particularly important in autonomous robotics because the target path to be traversed keeps changing dynamically with time. Hence, the trajectory controller for an autonomous robot has to be more robust and dynamic than that for a manually controlled robot \cite{manual}. For the robot design considered in this paper, the trajectory control faces even more challenges because of high level planning issues for an autonomous drive.
 \begin{figure}[htbp]
 			  \centering
 			  \def\svgscale{0.25}
  			  \tiny{
  			  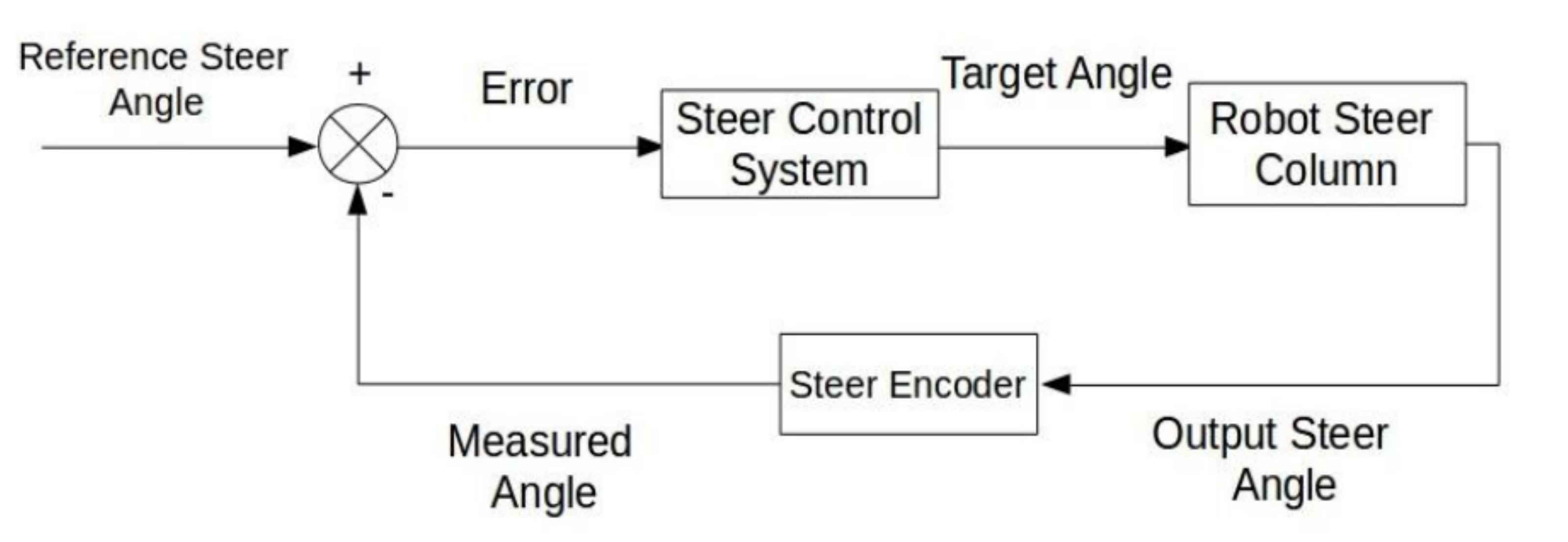}
 			  \caption{Steering Control System Block Diagram}
 			 \label{steering}
 		\end{figure}	
The trajectory control interacts with the steering control as shown in Fig.(\ref{trablock}). In this work our objective was to design the trajectory control strategy which feeds the steering control loop. The steering control loop has its own controller whose design is also considered in the paper. The steering control block diagram for the robot is shown in Fig.(\ref{steering}).
\section{Model Identification}
\label{sysid}
We used standard system identification techniques to identify the model of the three wheeled mobile robot with front steer. For the velocity control system, as mentioned above, we assumed that the robot dynamics are primarily due to the BLDC motor which is responsible for the translation. For a BLDC motor as shown in \cite{bldcmodel} and similar other works, we assumed a second order model with unknown parameters. By obtaining a set of input and corresponding output measurements, we used a system identification algorithm to obtain the unknown parameters. \\
The plant is a continuous time system, however, since the data from the sensors and the input to the plant are both discrete time, we used the sampling time for the data to identify a continuous time system model. A instrumental variable system identification approach was used to estimate the transfer function \cite{iv}.

\section{Control Design}
For response to a step input in velocity control, we designed a controller based on the model identified. A fast rise time is often the most desirable performance characteristic for any autonomous mobile robot. Other than the high bandwidth, the control design should be such that the closed loop system is insensitive to external disturbances which arise due to undulations in the road terrain and other environmental disturbances. To achieve both the design objectives a lead-lag compensator is needed. This lead-lag compensator has been implemented as a PID controller on a digital microcontroller platform. A digital control is not only very easy to implement compared to analog control, but also provides the option to change the reference input and controller parameters easily. \\
For the second order plant model identified $G(s)$, we designed a discrete-time controller $D(z)$ from the design specifications. We chose a desired rise time of $t_{r}$ for a step input velocity command to the robot and a phase lag to attenuate the disturbances at high frequency. In the transformed frequency domain, we can write the controller equation as follows (\cite{ogata}):
\begin{align}
D(w) &= K_{p} + \frac{K_{i}}{w} + K_{d}w\\
D(j\omega_{w_{1}}) &=  K_{p} + \frac{K_{i}}{j\omega_{w_{1}}} + K_{d}j\omega_{w_{1}}\\
\intertext{In polar representation}
D(j\omega_{w_{1}}) &= \left|D\right|\left(\cos(\theta) + \sin(\theta)\right)\\
\intertext{where $K_{p}$, $K_{i}$ and $K_{d}$ are ideal PID controller constants. At gain crossover frequency, $\omega_{w_{1}}$, we have}
\left|D\right| &= \frac{1}{\left|G\right|}\\
\intertext{where $\left|D\right|$ and $\left|G\right|$ are magnitude of the controller and the plant at $\omega_{w_{1}}$. For a given phase lag angle $\theta$, we can use the above equations to write the PID controller parameters as follows}
\label{Kpequation}
K_{p} &= \frac{\cos(\theta)}{\left|G\right|}\\
\label{Kdequation}
K_{d}\omega_{w_{1}} -\frac{K_{i}}{\omega_{w_{1}}} &= \frac{\sin(\theta)}{\left|G\right|}
\end{align}
Now, using Eq.(\ref{Kpequation}) and Eq.(\ref{Kdequation}), we can find the values of the PID controller parameters if we choose the value of one of them depending on the desired performance specifications. We used this control design methodology for the mobile robot shown in Fig.(\ref{robot}) and the results have been given in Section (\ref{controlimp}). 

\section{Trajectory control}
\label{trajectory}
Consider that the three wheeled mobile robot is traversing on a path with a curvature $\kappa$. The curvature of the path is defined as the inverse of the instantaneous radius of curvature, centered around a hypothetical center of a circle. The center of curvature is similarly defined as the center of a circle which passes through the path at a given point which has the same tangent and curvature at that point on the path.
We can calculate the curvature for the robot design in consideration in this paper as shown below. (Refer Fig.(\ref{trajgeo})).
\begin{figure}[htbp]

			  \centering
			  \def\svgscale{0.65}
			  \tiny{
			  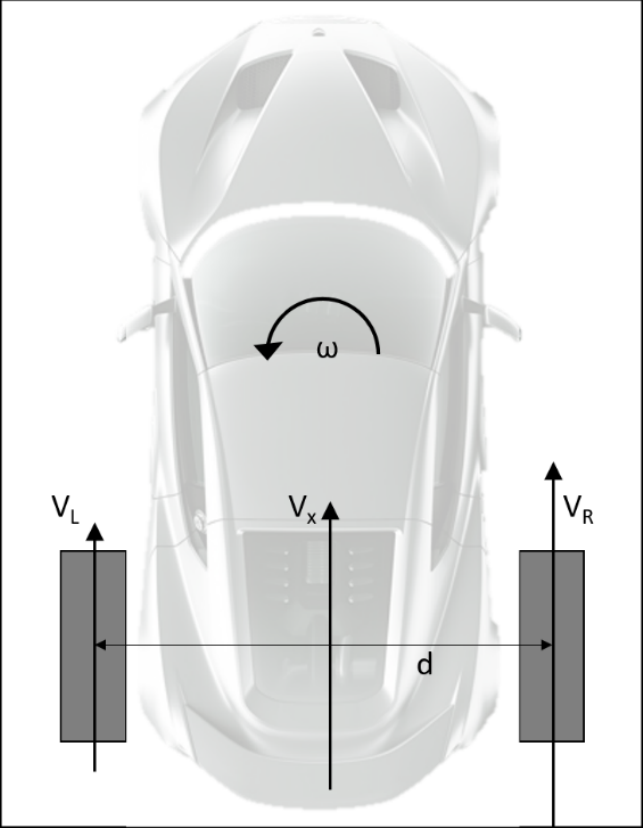}
			  \caption{Differential drive geometry}
			 \label{trajgeo}
		\end{figure}
\begin{align}
\label{5}
V_{L}&=r\times\omega_{L}\\
\label{6}
V_{R}&=r\times\omega_{R}
\end{align}
where $r$ is the radius of a wheel, $\omega_{L}$ is the left wheel angular velocity and $\omega_{R}$ is the right wheel angular velocity.\\
The rear wheels follow differential drive kinematics, as they are both free to move in both clockwise and anticlockwise directions. The simplistic differential drive model can be used to calculate the kinematic equations of the robot.
\begin{align}
\label{9}
V_{L}&=V_{x}-\frac d 2 \times\omega \\
\label{10}
V_{R}&=V_{x}+\frac d 2 \times\omega\\
\intertext{where $d$ is the separation between the two wheels, $\omega$ is the instantaneous angular velocity of the robot, assumed anticlockwise about a point midway between the wheels. Adding Eq.(\ref{9}) and Eq.(\ref{10}), we get}
V_{x}&=\frac{V_{L}+V_{R}}{2}\\
\intertext{Subtracting Eq.(\ref{10}) from Eq.(\ref{9}), we get}
\omega &= \frac {V_{L}-V_{R}}{d}
\end{align}
\subsection{Curvature estimation}
By definition, we can write the curvature as follows
\begin{equation}
\kappa=\frac 1 R
\end{equation}
where $\kappa$ is the curvature of the path and $R$ is the instantaneous radius of curvature. \\
Assuming the robot to be a rigid body, we can write
\begin{align}
\frac {V_{L}} {R-\frac d 2} &= \frac {V_{R}} {R+\frac d 2}\\
\intertext{Cross multiplying and solving, we get,}
\frac {V_{R}+V_{L}} {V_{R}-V_{L}} &= \frac {2R} {d}\\
\intertext{Rearranging terms, we get}
\frac {V_{L}+V_{R}}{2} \times \frac{1}{\frac{V_{R}-V_{L}}{d} } &= R\\
\intertext{Using equations (\ref{5}) and (\ref{6}) we can write}
\label{curve}
\kappa &= \frac{\omega}{V_{x}}
\end{align}

Using the relation in Eq.(\ref{curve}), the curvature control was designed. It was implemented as a separate PID control loop in the digital controller as shown in the Fig.(\ref{trablock}). The steering angle of the robot is controlled using the trajectory controller. \\
\begin{figure}[htbp]

			  \centering
			  \def\svgscale{0.25}
			  \tiny{
			  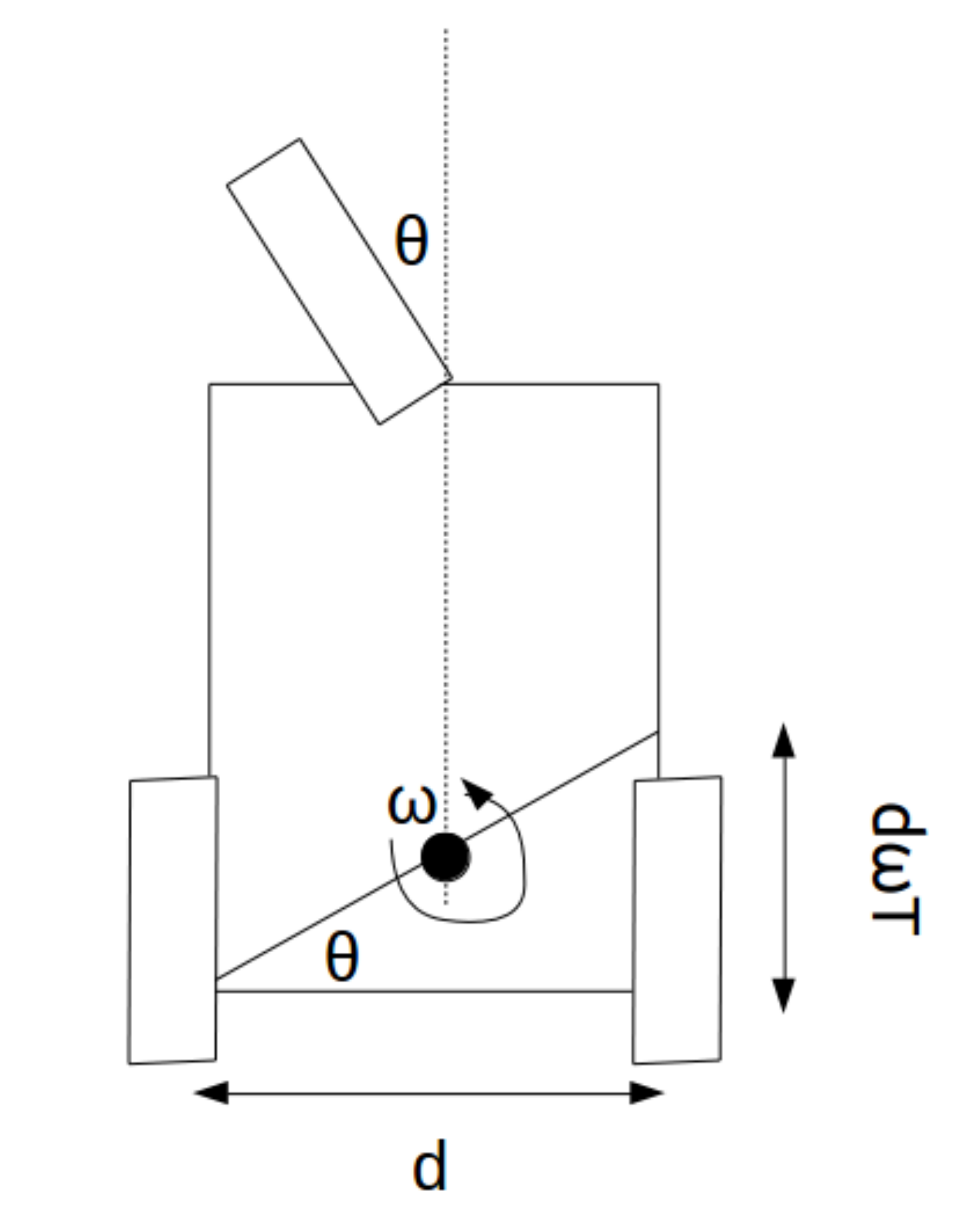}
			  \caption{Robot Geometry for Trajectory Control}
			 \label{figuretrig}
		\end{figure}
if the robot turns to its left at a constant curvature $\kappa$ for a time T, then analyzing the motion relative to the left wheel (considering it to be stationary), we can write that the right wheel moves a distance of $\omega \times d \times T$, where $\omega$ is the instantaneous angular velocity of the robot. From the Fig (\ref{figuretrig}), we can write,
\begin{equation}
\tan(\theta)= \omega \times T
\end{equation}
Hence, the steer angle is proportional to $\omega$ for small values of $\theta$. Using this linearization about small values of $\theta$, we can design the PID controller for the trajectory. Also, since $\kappa$ is proportional to $\omega$ for a constant velocity, the output of the trajectory controller would be the reference steering input angle to the steering angle control loop.\\
The controller parameters were experimentally tuned. The same module was also used to provide data for several other tasks not directly linked to low-level control, such as localization and Simultaneous Localization And Mapping (SLAM) for a successful autonomous run.

\section{Model Validation}
Using the system identification method described in Section (\ref{sysid}), the following continuous transfer function model was obtained for the robot velocity control system. The transfer function model was obtained incorporating the delay in the system as well. 
\begin{equation}
\label{tf}
G(s) = \frac{K(s+2.8)}{(s+0.44)(s+5)} \times \exp(-0.3s)
\end{equation}
The identified model was validated using comparisons with the experimental response of the robot to different inputs. The transfer function given in Eq.(\ref{tf}) is when the BLDC motor is running under the load of the whole robot on a road. A similar model was also identified by running the BLDC motor without load. The BLDC model without load was obtained to be consistent with the robot model for velocity control justifying our previous assumption that for velocity control, the system dynamics are majorly governed by the BLDC motor alone, which provides the translation torque to the robot. To obtain the identified transfer function as shown in Eq.(\ref{tf}), the robot was excited with a step input around a nominal operating point of $1m/s$ speed. A voltage input step command to the BLDC motor was given and the transient output response of the velocity was recorded. For the operating point $(11V, 1m/s)$, the linearized model was obtained. (11 V is the averaged voltage value when the BLDC is given a $21\%$ duty cycle input to its source voltage of 48 V). In the next section, we demonstrate the validity of the linearity assumption and the range of speeds for which the obtained model is valid. 
\subsection{Linearity}
\begin{figure}[htbp]

			  \centering
			  \def\svgscale{0.25}
			  \tiny{
			  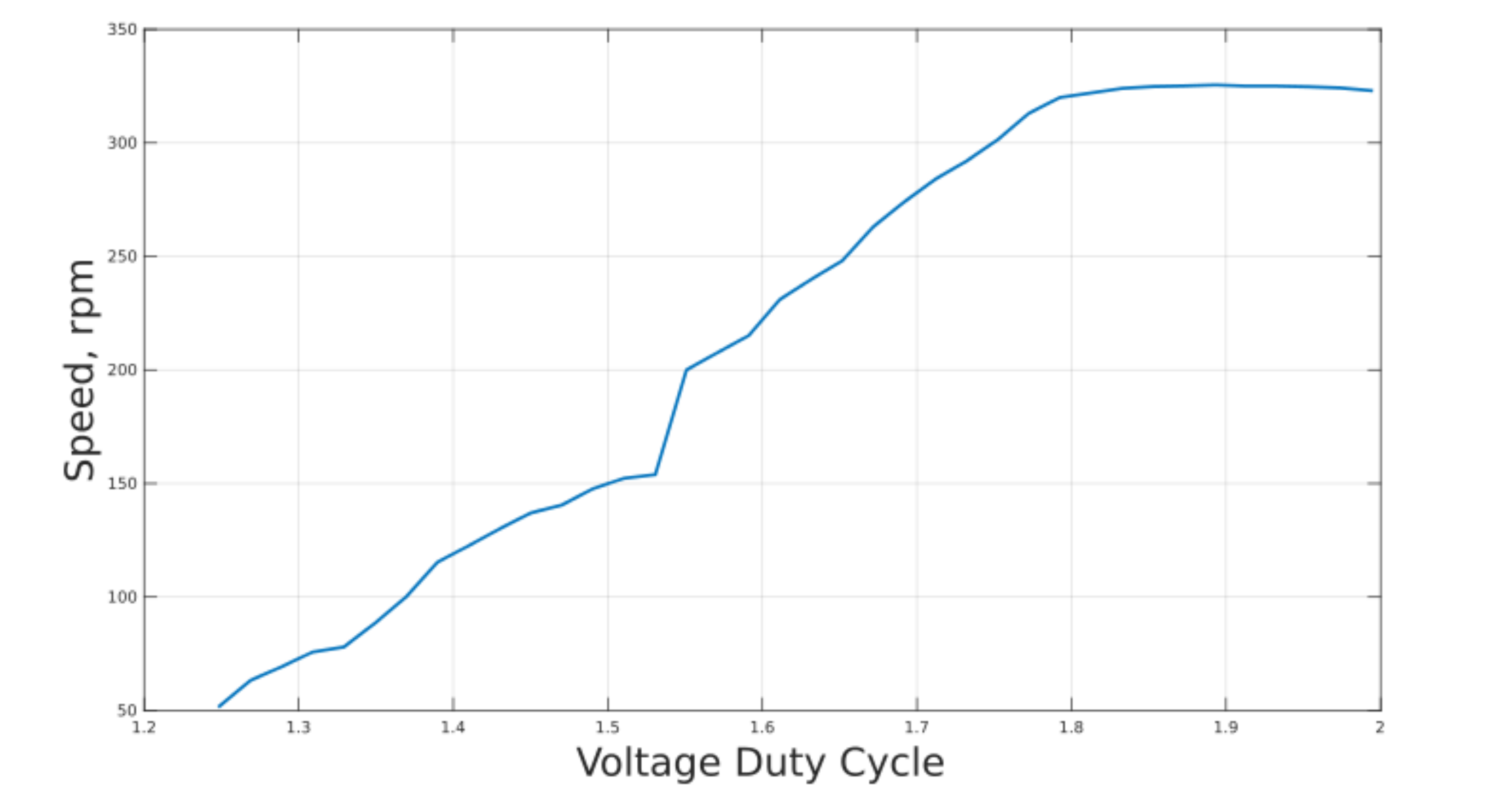}
			  \caption{Nonlinear voltage duty cycle and speed characteristics}
			 \label{bldcNL}
		\end{figure}
For the BLDC motor, the voltage-speed response was obtained experimentally. The resulting motor characteristic is shown in Fig.(\ref{bldcNL}). Clearly, the motor has nonlinear dynamics as the speed saturates after a certain limit. To obtain the linear model given in Eq.(\ref{tf}), linearization was done around the nominal speed of $1m/s$. Since, the model is being used to design the controller which works for the system at different speeds it is important to identify the range for which the robot behaves linearly, which would in effect give the range for which the designed controller would work.

\begin{figure}[htbp]

			  \centering
			  \def\svgscale{0.25}
			  \tiny{
			  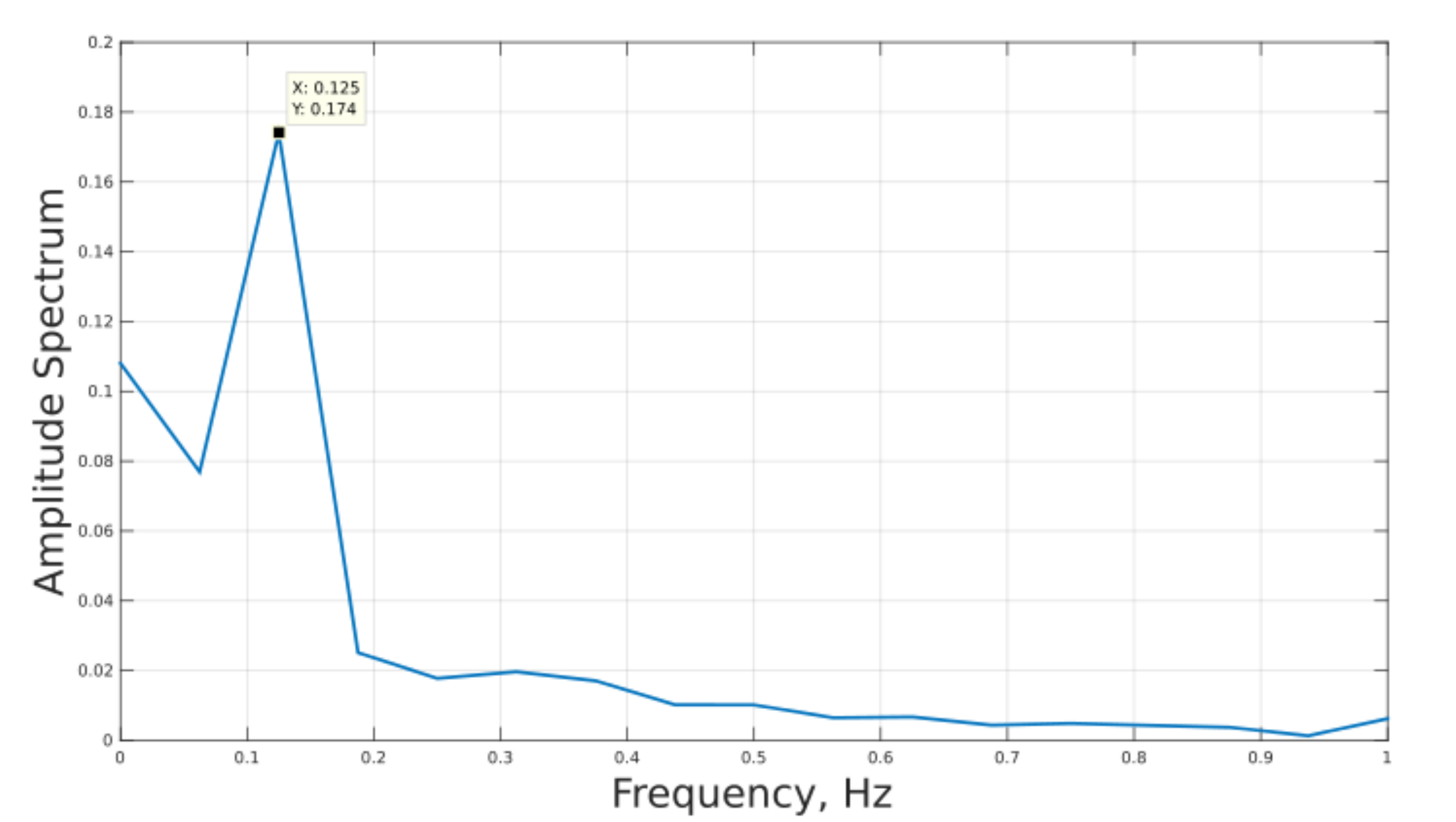}
			  \caption{Linearity Validation using Fourier analysis, the maximum frequency component corresponds to 0.125 Hz - the input frequency}
			 \label{sine}
		\end{figure}
To verify the superposition theorem to check the linearity range, the robot was excited with a sinusoidal input signal and the output was recorded. Using Fourier transform, the power density of each frequency component of the output response was obtained. This procedure was repeated for different amplitudes of the sinusoidal input signal around the operating point. From this frequency domain analysis, we observed that the linear model is valid for average voltage amplitudes of up to 28V, i.e. a  10V increment about the nominal operating voltage of 18 V. (These are average voltage values, the source voltage is a constant 48 V, under PWM changing duty cycle). The input signal frequency given was 0.125 Hz. The Fig.(\ref{sine}) shows the frequency component amplitudes in the output response. Clearly, the 0.125 Hz frequency has the maximum power, proving the fact that for this increment the linear model holds. This increment in voltage corresponds to $4m/s$ velocity. Hence, we operate our designed controller in this velocity range.    

\subsection{Open Loop Performance}
\begin{figure}[htbp]

			  \centering
			  \def\svgscale{0.25}
			  \tiny{
			  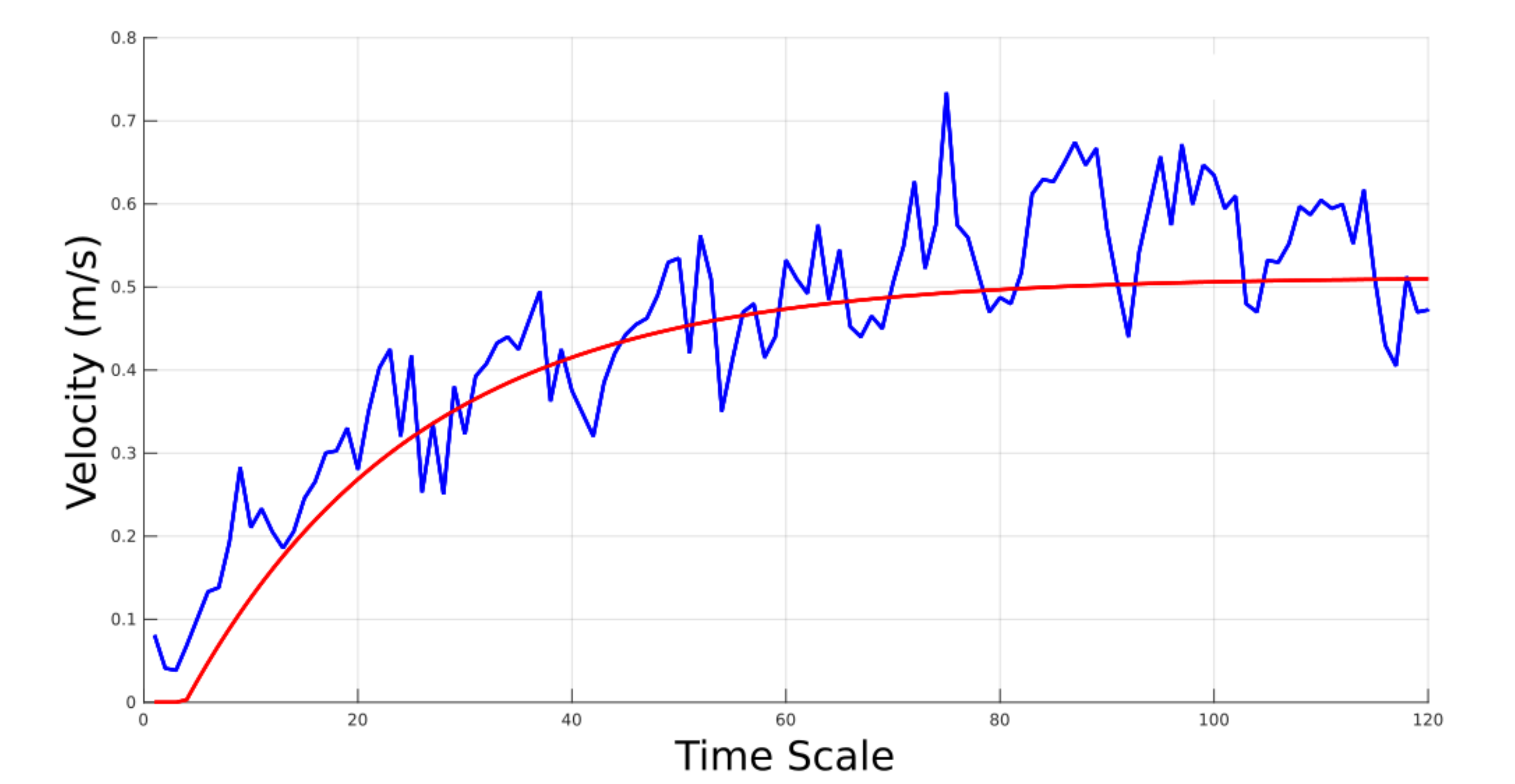}
			  \caption{Model validation - Step input of amplitude 5 V (10 \% duty cycle) above 11 V operating point. Red - Identified model result, Blue - Experimental result}
			 \label{openloop1}
		\end{figure}

\begin{figure}[htbp]

			  \centering
			  \def\svgscale{0.25}
			  \tiny{
			  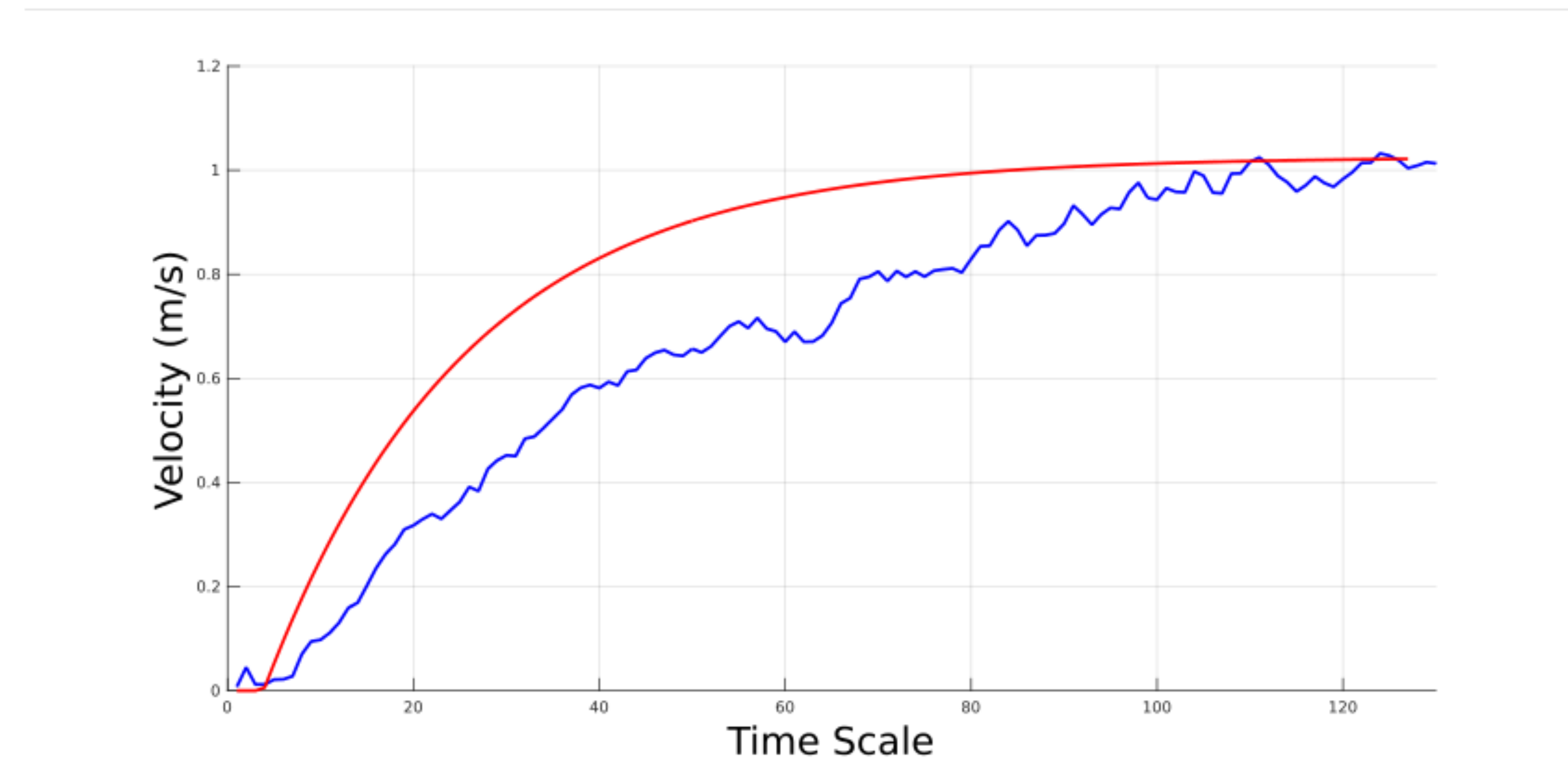}
			  \caption{Model validation - Step input of amplitude 10 V (20 \% duty cycle) above 11 V operating point. Red - Identified model result, Blue - Experimental result}
			 \label{openloop2}
		\end{figure}

\begin{figure}[htbp]

			  \centering
			  \def\svgscale{0.25}
			  \tiny{
			  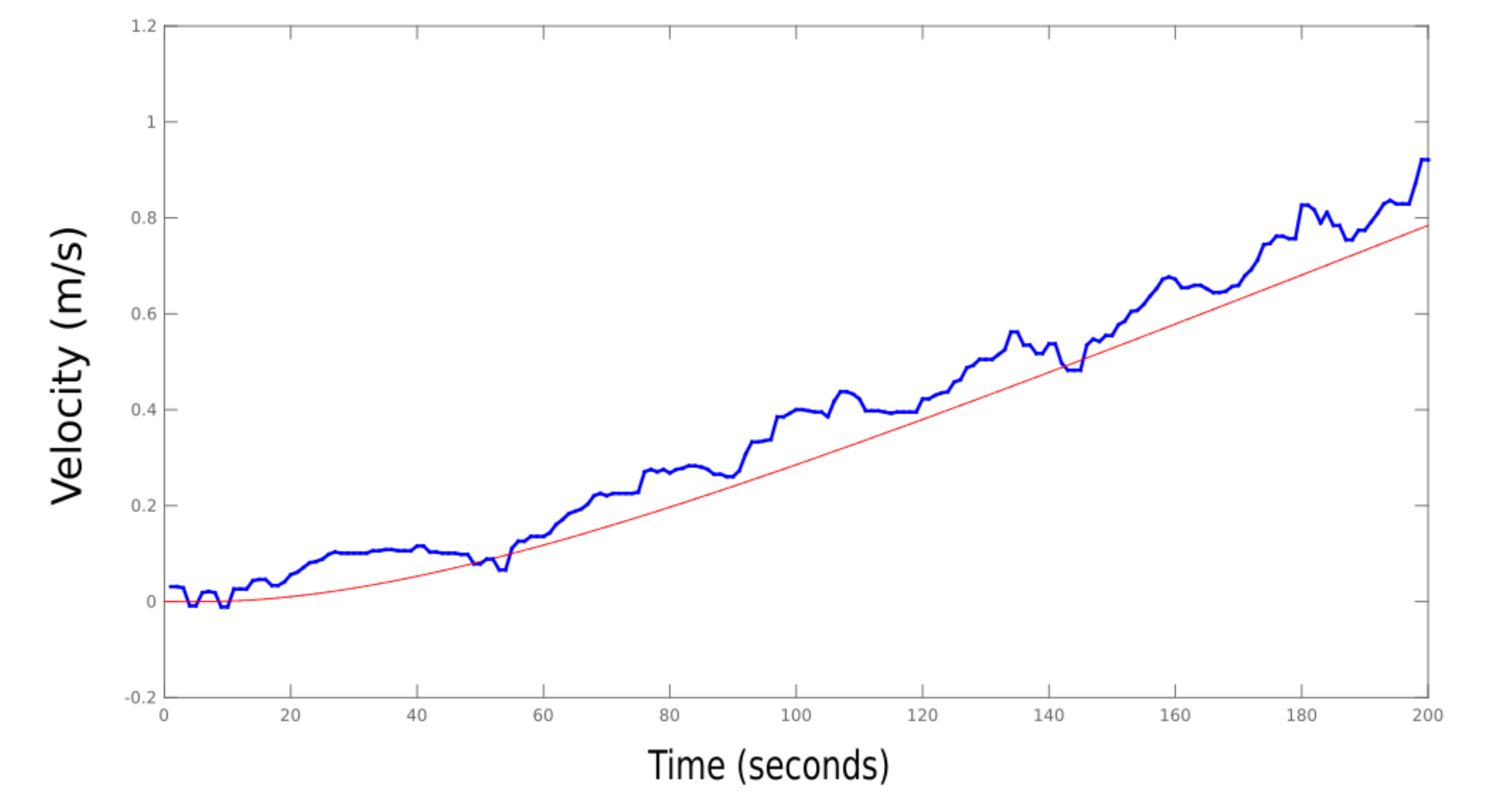}
			  \caption{Model validation - Ramp input. Red - Identified model result, Blue - Experimental result}
			 \label{openloop3}
		\end{figure}

To validate the robot model obtained for velocity control system, we compared the open loop performance of the robot using experimental results and the response as calculated from the model. A comparison is shown in Fig.(\ref{openloop1}),Fig.(\ref{openloop2}) and Fig.(\ref{openloop3}) for two step inputs of different amplitudes and a ramp input. 
	

The open loop performance of the robot was compared to the model response for other test input signals as well such as a ramp input signal to validate the model obtained. 

\section{Control design and implementation}
\label{controlimp}
\begin{figure}[htbp]

			  \centering
			  \def\svgscale{0.25}
			  \tiny{
			  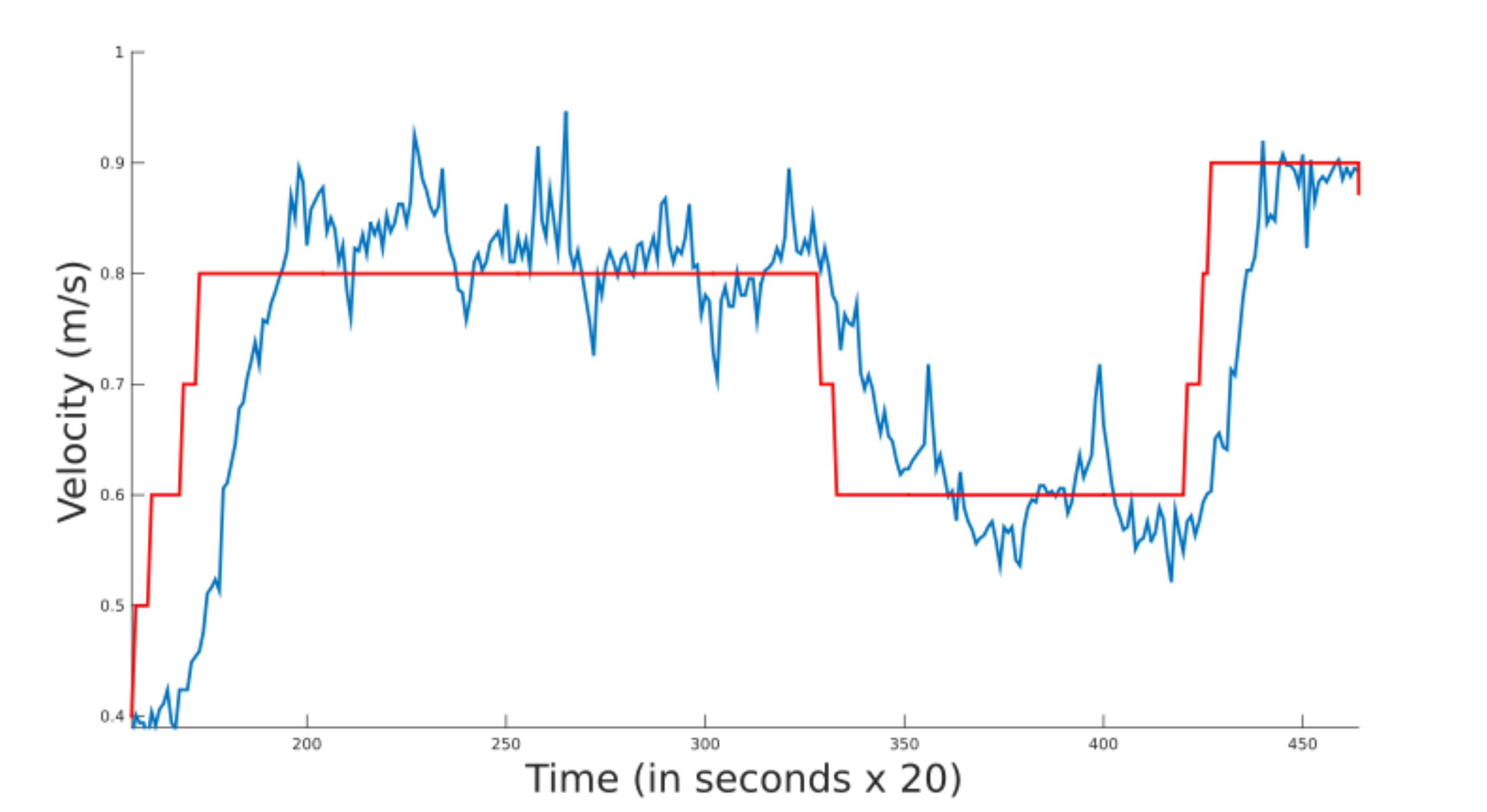}
			  \caption{Closed loop performance of the robot using the designed controller}
			 \label{closedloop}
		\end{figure}

A PID controller was designed for velocity control based on the identified system model. To achieve a rise time of 0.5 seconds and a phase lag of $5^{\circ}$ to attenuate the high frequency noise, we used equations (\ref{Kpequation} and \ref{Kdequation}) to calculate the values for the controller parameters. On choosing the value of $K_{i}$ to achieve the desired lag response, we calculated the values for $K_{p}$  and $K_{d}$ using the equations given above. The discrete-time controller can then be written as shown below.
\begin{equation}
D(z) = K_{p} + \frac{K_{i} T }{2}\frac{z+1}{z-1} + \frac{K_{d} (z-1)}{Tz}
\end{equation}
The discrete time controller equation was obtained by using bilinear transformation from the transform domain to z-domain for the integrator and the backward difference method for the differentiator to incorporate the finite bandwidth differentiator in the controller. The sampling time for the implemented controller was 0.05 seconds. Using the PID parameter values and the sampling time, the controller was implemented on a computer (Intel 64 bit microprocessor) in the discrete-time domain. The control input was given to a DAC which provides the input to the BLDC motor in terms of the duty cycle according to the given control input. For system analysis, the effect of this DAC was incorporated by obtaining a zero-order hold equivalent of the continuous-time plant. The closed loop performance was then analyzed in discrete-time domain. The response of the closed loop system to a step input was similar to the experimental closed loop response. The robot performance is shown in Fig.(\ref{closedloop}).


\section{Experimental Results - Trajectory Control}
Apart from velocity control, the trajectory control of the three wheeled autonomous mobile robot with front steer poses newer challenges which are previously uncovered in the existing literature. This was described in detail in Section (\ref{trajectory}). The trajectory controller was designed for the robot using the method given in curvature estimation section. The controller was implemented in the structure as shown in Fig.(\ref{trablock}) and the performance of the robot was analyzed.

\section{Future Work}
The work on autonomous three wheeled robots has a long way to go before such vehicles are realized onto the roads because of many control and stability related challenges. We presented the modeling and control design approach for a particular kind of three wheeled robots. This work could be carried forward in various directions such as design of controllers using the robust and optimal control theoretic techniques so that the robot performs under various uncertainties while consuming as less battery power as possible. It would also be interesting to design adaptive PID controllers for which the parameters change according to the environment conditions. Although, there has been a significant amount of research on adaptive and fuzzy PID control designs, but extending such results to this robot design would be an interesting problem to consider given the different challenges that this front wheel driven \& steered design poses.  

\section{Conclusion}
We identified a linearized model for a three wheeled autonomous mobile robot. The robot design considered in the paper is a front wheel steer design. The model was validated by comparing the derived model response with the experimental results of the autonomous robot. The linearity of the model was also investigated thoroughly and the range of linearity was calculated by analyzing the experimental data. A PID controller was designed based on the identified model and was implemented in a discrete-time controller hardware. The trajectory control problem was touched upon briefly in this paper and some promising initial results were demonstrated. A high level planner designed for holonomic differential drive robots only, when used with the designed trajectory controller for the three wheeled robot design followed the desired trajectory.


\section*{Acknowledgment}

The authors would like to thank all the members of the Autonomous Ground Vehicle (AGV) research group, IIT Kharagpur for their continuous support in our research work. We would like to extend special gratitude towards Jignesh Sindha (PhD research scholar in the group) for his help with the robot mechanical design. We would also like to thank Gopabandhu Hota, Yash Gaurkar and Aakash Yadav for letting us use their velocity measurement sensor to obtain the experimental results. We are grateful to Sponsored Research and Industrial Consultancy (SRIC), IIT Kharagpur for funding the research in our group.



%
\printbibliography

\end{document}

%% file: robot.pdf_tex
\begingroup%
  \makeatletter%
  \providecommand\color[2][]{%
    \errmessage{(Inkscape) Color is used for the text in Inkscape, but the package 'color.sty' is not loaded}%
    \renewcommand\color[2][]{}%
  }%
  \providecommand\transparent[1]{%
    \errmessage{(Inkscape) Transparency is used (non-zero) for the text in Inkscape, but the package 'transparent.sty' is not loaded}%
    \renewcommand\transparent[1]{}%
  }%
  \providecommand\rotatebox[2]{#2}%
  \ifx\svgwidth\undefined%
    \setlength{\unitlength}{829.00003268bp}%
    \ifx\svgscale\undefined%
      \relax%
    \else%
      \setlength{\unitlength}{\unitlength * \real{\svgscale}}%
    \fi%
  \else%
    \setlength{\unitlength}{\svgwidth}%
  \fi%
  \global\let\svgwidth\undefined%
  \global\let\svgscale\undefined%
  \makeatother%
  \begin{picture}(1,0.56574182)%
    \put(0,0){\includegraphics[width=\unitlength,page=1]{robot.pdf}}%
  \end{picture}%
\endgroup%

%% file: velocity_control.pdf_tex
\begingroup%
  \makeatletter%
  \providecommand\color[2][]{%
    \errmessage{(Inkscape) Color is used for the text in Inkscape, but the package 'color.sty' is not loaded}%
    \renewcommand\color[2][]{}%
  }%
  \providecommand\transparent[1]{%
    \errmessage{(Inkscape) Transparency is used (non-zero) for the text in Inkscape, but the package 'transparent.sty' is not loaded}%
    \renewcommand\transparent[1]{}%
  }%
  \providecommand\rotatebox[2]{#2}%
  \ifx\svgwidth\undefined%
    \setlength{\unitlength}{870.99998654bp}%
    \ifx\svgscale\undefined%
      \relax%
    \else%
      \setlength{\unitlength}{\unitlength * \real{\svgscale}}%
    \fi%
  \else%
    \setlength{\unitlength}{\svgwidth}%
  \fi%
  \global\let\svgwidth\undefined%
  \global\let\svgscale\undefined%
  \makeatother%
  \begin{picture}(1,0.37428241)%
    \put(0,0){\includegraphics[width=\unitlength,page=1]{velocity_control.pdf}}%
  \end{picture}%
\endgroup%

%% file: steering_control.pdf_tex
\begingroup%
  \makeatletter%
  \providecommand\color[2][]{%
    \errmessage{(Inkscape) Color is used for the text in Inkscape, but the package 'color.sty' is not loaded}%
    \renewcommand\color[2][]{}%
  }%
  \providecommand\transparent[1]{%
    \errmessage{(Inkscape) Transparency is used (non-zero) for the text in Inkscape, but the package 'transparent.sty' is not loaded}%
    \renewcommand\transparent[1]{}%
  }%
  \providecommand\rotatebox[2]{#2}%
  \ifx\svgwidth\undefined%
    \setlength{\unitlength}{920.99997693bp}%
    \ifx\svgscale\undefined%
      \relax%
    \else%
      \setlength{\unitlength}{\unitlength * \real{\svgscale}}%
    \fi%
  \else%
    \setlength{\unitlength}{\svgwidth}%
  \fi%
  \global\let\svgwidth\undefined%
  \global\let\svgscale\undefined%
  \makeatother%
  \begin{picture}(1,0.34527686)%
    \put(0,0){\includegraphics[width=\unitlength,page=1]{steering_control.pdf}}%
  \end{picture}%
\endgroup%

%% file: trajblock.pdf_tex
\begingroup%
  \makeatletter%
  \providecommand\color[2][]{%
    \errmessage{(Inkscape) Color is used for the text in Inkscape, but the package 'color.sty' is not loaded}%
    \renewcommand\color[2][]{}%
  }%
  \providecommand\transparent[1]{%
    \errmessage{(Inkscape) Transparency is used (non-zero) for the text in Inkscape, but the package 'transparent.sty' is not loaded}%
    \renewcommand\transparent[1]{}%
  }%
  \providecommand\rotatebox[2]{#2}%
  \ifx\svgwidth\undefined%
    \setlength{\unitlength}{1020.00002884bp}%
    \ifx\svgscale\undefined%
      \relax%
    \else%
      \setlength{\unitlength}{\unitlength * \real{\svgscale}}%
    \fi%
  \else%
    \setlength{\unitlength}{\svgwidth}%
  \fi%
  \global\let\svgwidth\undefined%
  \global\let\svgscale\undefined%
  \makeatother%
  \begin{picture}(1,0.4254902)%
    \put(0,0){\includegraphics[width=\unitlength,page=1]{trajblock.pdf}}%
  \end{picture}%
\endgroup%

%% file: differential.pdf_tex
\begingroup%
  \makeatletter%
  \providecommand\color[2][]{%
    \errmessage{(Inkscape) Color is used for the text in Inkscape, but the package 'color.sty' is not loaded}%
    \renewcommand\color[2][]{}%
  }%
  \providecommand\transparent[1]{%
    \errmessage{(Inkscape) Transparency is used (non-zero) for the text in Inkscape, but the package 'transparent.sty' is not loaded}%
    \renewcommand\transparent[1]{}%
  }%
  \providecommand\rotatebox[2]{#2}%
  \ifx\svgwidth\undefined%
    \setlength{\unitlength}{184.90908203bp}%
    \ifx\svgscale\undefined%
      \relax%
    \else%
      \setlength{\unitlength}{\unitlength * \real{\svgscale}}%
    \fi%
  \else%
    \setlength{\unitlength}{\svgwidth}%
  \fi%
  \global\let\svgwidth\undefined%
  \global\let\svgscale\undefined%
  \makeatother%
  \begin{picture}(1,1.28849561)%
    \put(0,0){\includegraphics[width=\unitlength,page=1]{differential.pdf}}%
  \end{picture}%
\endgroup%

%% file: figuretrig.pdf_tex
\begingroup%
  \makeatletter%
  \providecommand\color[2][]{%
    \errmessage{(Inkscape) Color is used for the text in Inkscape, but the package 'color.sty' is not loaded}%
    \renewcommand\color[2][]{}%
  }%
  \providecommand\transparent[1]{%
    \errmessage{(Inkscape) Transparency is used (non-zero) for the text in Inkscape, but the package 'transparent.sty' is not loaded}%
    \renewcommand\transparent[1]{}%
  }%
  \providecommand\rotatebox[2]{#2}%
  \ifx\svgwidth\undefined%
    \setlength{\unitlength}{423.0000173bp}%
    \ifx\svgscale\undefined%
      \relax%
    \else%
      \setlength{\unitlength}{\unitlength * \real{\svgscale}}%
    \fi%
  \else%
    \setlength{\unitlength}{\svgwidth}%
  \fi%
  \global\let\svgwidth\undefined%
  \global\let\svgscale\undefined%
  \makeatother%
  \begin{picture}(1,1.24586287)%
    \put(0,0){\includegraphics[width=\unitlength,page=1]{figuretrig.pdf}}%
  \end{picture}%
\endgroup%

%% file: bldcNL.pdf_tex
\begingroup%
  \makeatletter%
  \providecommand\color[2][]{%
    \errmessage{(Inkscape) Color is used for the text in Inkscape, but the package 'color.sty' is not loaded}%
    \renewcommand\color[2][]{}%
  }%
  \providecommand\transparent[1]{%
    \errmessage{(Inkscape) Transparency is used (non-zero) for the text in Inkscape, but the package 'transparent.sty' is not loaded}%
    \renewcommand\transparent[1]{}%
  }%
  \providecommand\rotatebox[2]{#2}%
  \ifx\svgwidth\undefined%
    \setlength{\unitlength}{795.99995771bp}%
    \ifx\svgscale\undefined%
      \relax%
    \else%
      \setlength{\unitlength}{\unitlength * \real{\svgscale}}%
    \fi%
  \else%
    \setlength{\unitlength}{\svgwidth}%
  \fi%
  \global\let\svgwidth\undefined%
  \global\let\svgscale\undefined%
  \makeatother%
  \begin{picture}(1,0.52512564)%
    \put(0,0){\includegraphics[width=\unitlength,page=1]{bldcNL.pdf}}%
  \end{picture}%
\endgroup%

%% file: sine.pdf_tex
\begingroup%
  \makeatletter%
  \providecommand\color[2][]{%
    \errmessage{(Inkscape) Color is used for the text in Inkscape, but the package 'color.sty' is not loaded}%
    \renewcommand\color[2][]{}%
  }%
  \providecommand\transparent[1]{%
    \errmessage{(Inkscape) Transparency is used (non-zero) for the text in Inkscape, but the package 'transparent.sty' is not loaded}%
    \renewcommand\transparent[1]{}%
  }%
  \providecommand\rotatebox[2]{#2}%
  \ifx\svgwidth\undefined%
    \setlength{\unitlength}{748bp}%
    \ifx\svgscale\undefined%
      \relax%
    \else%
      \setlength{\unitlength}{\unitlength * \real{\svgscale}}%
    \fi%
  \else%
    \setlength{\unitlength}{\svgwidth}%
  \fi%
  \global\let\svgwidth\undefined%
  \global\let\svgscale\undefined%
  \makeatother%
  \begin{picture}(1,0.56417112)%
    \put(0,0){\includegraphics[width=\unitlength,page=1]{sine.pdf}}%
  \end{picture}%
\endgroup%

%% file: steplow.pdf_tex
\begingroup%
  \makeatletter%
  \providecommand\color[2][]{%
    \errmessage{(Inkscape) Color is used for the text in Inkscape, but the package 'color.sty' is not loaded}%
    \renewcommand\color[2][]{}%
  }%
  \providecommand\transparent[1]{%
    \errmessage{(Inkscape) Transparency is used (non-zero) for the text in Inkscape, but the package 'transparent.sty' is not loaded}%
    \renewcommand\transparent[1]{}%
  }%
  \providecommand\rotatebox[2]{#2}%
  \ifx\svgwidth\undefined%
    \setlength{\unitlength}{1086.00000577bp}%
    \ifx\svgscale\undefined%
      \relax%
    \else%
      \setlength{\unitlength}{\unitlength * \real{\svgscale}}%
    \fi%
  \else%
    \setlength{\unitlength}{\svgwidth}%
  \fi%
  \global\let\svgwidth\undefined%
  \global\let\svgscale\undefined%
  \makeatother%
  \begin{picture}(1,0.51749539)%
    \put(0,0){\includegraphics[width=\unitlength,page=1]{steplow.pdf}}%
  \end{picture}%
\endgroup%

%% file: step1.pdf_tex
\begingroup%
  \makeatletter%
  \providecommand\color[2][]{%
    \errmessage{(Inkscape) Color is used for the text in Inkscape, but the package 'color.sty' is not loaded}%
    \renewcommand\color[2][]{}%
  }%
  \providecommand\transparent[1]{%
    \errmessage{(Inkscape) Transparency is used (non-zero) for the text in Inkscape, but the package 'transparent.sty' is not loaded}%
    \renewcommand\transparent[1]{}%
  }%
  \providecommand\rotatebox[2]{#2}%
  \ifx\svgwidth\undefined%
    \setlength{\unitlength}{845.00001922bp}%
    \ifx\svgscale\undefined%
      \relax%
    \else%
      \setlength{\unitlength}{\unitlength * \real{\svgscale}}%
    \fi%
  \else%
    \setlength{\unitlength}{\svgwidth}%
  \fi%
  \global\let\svgwidth\undefined%
  \global\let\svgscale\undefined%
  \makeatother%
  \begin{picture}(1,0.49112423)%
    \put(0,0){\includegraphics[width=\unitlength,page=1]{step1.pdf}}%
  \end{picture}%
\endgroup%

%% file: ramp.pdf_tex
\begingroup%
  \makeatletter%
  \providecommand\color[2][]{%
    \errmessage{(Inkscape) Color is used for the text in Inkscape, but the package 'color.sty' is not loaded}%
    \renewcommand\color[2][]{}%
  }%
  \providecommand\transparent[1]{%
    \errmessage{(Inkscape) Transparency is used (non-zero) for the text in Inkscape, but the package 'transparent.sty' is not loaded}%
    \renewcommand\transparent[1]{}%
  }%
  \providecommand\rotatebox[2]{#2}%
  \ifx\svgwidth\undefined%
    \setlength{\unitlength}{1067.99998847bp}%
    \ifx\svgscale\undefined%
      \relax%
    \else%
      \setlength{\unitlength}{\unitlength * \real{\svgscale}}%
    \fi%
  \else%
    \setlength{\unitlength}{\svgwidth}%
  \fi%
  \global\let\svgwidth\undefined%
  \global\let\svgscale\undefined%
  \makeatother%
  \begin{picture}(1,0.5346442)%
    \put(0,0){\includegraphics[width=\unitlength,page=1]{ramp.pdf}}%
  \end{picture}%
\endgroup%

%% file: closedloop.pdf_tex
\begingroup%
  \makeatletter%
  \providecommand\color[2][]{%
    \errmessage{(Inkscape) Color is used for the text in Inkscape, but the package 'color.sty' is not loaded}%
    \renewcommand\color[2][]{}%
  }%
  \providecommand\transparent[1]{%
    \errmessage{(Inkscape) Transparency is used (non-zero) for the text in Inkscape, but the package 'transparent.sty' is not loaded}%
    \renewcommand\transparent[1]{}%
  }%
  \providecommand\rotatebox[2]{#2}%
  \ifx\svgwidth\undefined%
    \setlength{\unitlength}{789bp}%
    \ifx\svgscale\undefined%
      \relax%
    \else%
      \setlength{\unitlength}{\unitlength * \real{\svgscale}}%
    \fi%
  \else%
    \setlength{\unitlength}{\svgwidth}%
  \fi%
  \global\let\svgwidth\undefined%
  \global\let\svgscale\undefined%
  \makeatother%
  \begin{picture}(1,0.54626109)%
    \put(0,0){\includegraphics[width=\unitlength,page=1]{closedloop.pdf}}%
  \end{picture}%
\endgroup%